\documentstyle[aps,prd,epsf]{revtex}

\newcommand{\be}{\begin{equation}}
\newcommand{\ee}{\end{equation}}
\newcommand{\ben}{\begin{eqnarray}
\displaystyle}
\newcommand{\een}{\end{eqnarray}}

\newcommand{\la}{{\lambda}}

\newcommand{\p}{\partial}
\newcommand{\na}{\nabla}

\newcommand{\Lie}{{\cal L}}

\newcommand{\ga}{\gamma}

\begin{document}

\title{Uniqueness Theorem for Stationary Black Hole Solutions of $\sigma$-models
 in Five-dimensions}

\author{Marek Rogatko}

\address{Institute of Physics \protect \\
Maria Curie-Sklodowska University \protect \\
20-031 Lublin, pl.Marii Curie-Sklodowskiej 1, Poland \protect \\
rogat@tytan.umcs.lublin.pl \protect \\
rogat@kft.umcs.lublin.pl}

\date{\today}

\maketitle

\begin{abstract}
We prove the uniqueness theorem for stationary self-gravitating non-linear 
$\sigma$-models in five-dimensional spacetime. We show that the Myers-Perry 
vacuum Kerr spacetime is the only maximally extended, stationary, axisymmetric
asymptotically flat solution having the regular rotating event horizon with
a constant mapping.
\end{abstract}


\baselineskip=18pt
\section{Introduction}
Recently interests in higher dimensional black holes have renewed.
The unification attempts such as M/string theory
described our Universe
as a brane or defect emerged in higher dimensional geometry. 
$E8 \times E8$ heterotic string theory
at strong coupling is described in terms of M-theory acting in eleven-dimensional
spacetime with boundaries where ten-dimensional Yang-Mills gauge theories reside on two
boundaries \cite{hor}. 
The so-called
TeV gravity attracts attention to higher dimensional black hole 
because of the suggestion that such kind of objects may be 
produced in the near future, in high energy experiments \cite{gid}.
This kind of black holes are classical solutions of higher dimensional 
Einstein's equations. The radius of their event horizon is much smaller
that the scale of extra dimensions.
\par
The classification of non-singular black hole solutions
began with the Israel's work
\cite{isr}, then
M\"uller zum Hagen {\it et al.} \cite{mil73} and Robinson \cite{rob77},
provided other
contributions to the problem \cite{bun87,ru,ma1,he1,he93}.
In Refs.\cite{chr99a,chr99b} both for vacuum and Einstein-Maxwell (EM) black holes
the condition of non-degeneracy of the 
event horizon was removed. It was shown that for the static
electro-vacuum black holes all degenerate components of 
the event horizon should have charges of the same signs.
\par
On the other hand 
the problem of uniqueness theorem for stationary axisymmetric black hole
was considered in Refs.\cite{car,rob} and the complete proof was delivered by
Mazur \cite{maz} and Bunting \cite{bun}
(see also for a review of the uniqueness of black hole
solutions story see \cite{book} and references therein).
\par
Studies of the low-energy string theory 
also triggers the 
resurgence of works concerning the mathematical aspects of the
black holes emerging in it. Namely,
the staticity theorem for Einstein-Maxwell axion dilaton (EMAD) gravity
was studied in Ref.\cite{sta} and
uniqueness of the black hole solutions in dilaton gravity was  proved in 
works
\cite{dil,mar01}, while the uniqueness of the static
dilaton $U(1)^2$
black holes being the solution of $N = 4, d = 4$ supergravity
was provided in \cite{rog99}. The extension of the proof
to
$U(1)^{N}$ static dilaton black holes was established in Ref.\cite{rog02}.
\par
The possibility of production of higher dimensional black holes in 
accelerators caused 
the considerable interests in
$n$-dimensional black hole uniqueness theorem, both in vacuum and
charged case \cite{gib03,gib02a,gib02b,kod04}. The complete 
classification of $n$-dimensional
charged black holes having both degenerate and non-degenerate
components of event horizon was provided in Ref.\cite{rog03}.
Uniqueness theorem for $n$-dimensional static black hole carrying
{\it electric} and {\it magnetic} components of $(n-2)$-gauge
form was proved in \cite{rog04}.
\par
The primary signature of appearing black holes in future
accelerator experiments will be a Hawking emission and generically
the black holes will have angular momenta. Therefore the uniqueness
of rotating black holes in higher dimensions is of a great importance in
studies mathematical properties of them.
However the problem of 
uniqueness theorem for stationary 
$n$-dimensional black holes is much more complicated.
It was shown that generalization of Kerr metric to arbitrary $n$-dimensions proposed by
Myers-Perry \cite{mye86} is not unique (see for the counterexample showing that a five-dimensional
rotating black hole ring solution has the same angular momentum and mass
but its horizon is homeomorphic to $S^{2} \times S^{1}$ Refs.
\cite{emp02} and \cite{emp04}). It has been proved that
Myers-Perry solution is the unique black hole in five-dimensions in the class of 
spherical topology
and three commuting Killing vectors \cite{mor04}. The conditions for uniqueness of black holes
or black rings in
five-dimensional gravity was discussed in Ref.\cite{tom04}.
\par
The uniqueness theorem for self-gravitating nonlinear $\sigma$-models
in higher dimensional spacetime was obtained in \cite{rog02a}.
Recently, the related issues 
for supersymmetric black holes were given in Refs.\cite{sup}.
\par
In this paper we shall treat the problem of uniqueness of five-dimensional
axisymmetric, stationary self-gravitating $\sigma$-models. The main result established
in our work is that the only regular black hole solution with regular rotating event horizon
is the five-dimensional vacuum Kerr black hole with constant mapping.

\section{Five-dimensional rotating $\sigma$-models}
In our paper we shall consider
the action describing the $n$-dimensional self-gravitating $\sigma$-model
written as
\be
I = \int d^{n}x~\sqrt{- {}^{(n)} g} \bigg[
{}^{(n)}R - {1\over 2} G_{AB}(\varphi (x))~ \varphi^{A}_{, \mu}
\varphi^{B ,\mu} \bigg].
\ee
It then follows directly that
the energy momentum for the model has the form
\be
T_{\mu \nu}(\varphi) = G_{AB}(\varphi (x))~ \varphi^{A}_{,\mu} 
\varphi^{B}_{,\nu} - {1 \over 2} G_{AB}(\varphi (x))~ 
\varphi^{A}_{,\ga} \varphi^{B ,\ga} g_{\mu \nu}.
\ee
The equations of motion derived from the variational principle are as
\ben
\na_{\ga}\na^{\ga} \varphi^{A} &+& \Gamma^{A}_{BC}~
\varphi^{B}_{,\mu} \varphi^{C ,\mu} = 0,\\
{}^{(n)}G_{\mu \nu} &=& T_{\mu \nu}(\varphi).
\een
In what follows we shall take into account the asymptotically, five-dimensional
flat spacetime, i.e., the spacetime will contain a data set 
$(\Sigma_{end}, g_{ij}, K_{ij})$ with scalar fields of $\varphi$
such that a spacelike hypersurface $\Sigma_{end}$ is diffeomorphic to $\bf R^{4}$ minus 
a ball. The asymptotical conditions of the following forms should also be satisfied:
\be
\mid g_{ij} - \delta_{ij} \mid +
r \mid \p_{a} g_{ij} \mid + \dots
+r^{m} \mid \p_{a_{1} \dots a_{m}} g_{ij} \mid + r \mid K_{ij}\mid + \dots
+r^{m} \mid \p_{a_{1} \dots a_{m-1}} K_{ij} \mid \le {\cal O}\bigg( {1 \over r} \bigg),
\ee
where $g_{ij}$ and $K_{ij}$ are induced on $\Sigma_{end}$. $K_{ij}$ is the extrinsic
curvature tensor of the hypersurface $\Sigma_{end}$. 
It is required
that in the local coordinates on $\Sigma_{end}$ the scalar
field satisfies the following fall-off condition:
\be
\varphi^{A} =
\varphi^{A}_{\infty} + {\cal O}\bigg( {1 \over r^{3/2}} \bigg),
\ee
Our spacetime will admit three commutating Killing vector 
fields $k_{\mu}, \phi_{\mu}, \psi_{\mu}$
\be
[k, \phi] = [k, \psi] = [\phi, \psi] = 0,
\ee
where $k_{\mu}$ is an asymptotically timelike for which $V = - k_{\mu}k^{\mu}$, while
vectors $\phi_{\mu}$ and $\psi_{\mu}$ are spacelike. They all have closed orbits and
the following is satisfied:
\be
\Lie_{k}~g_{\mu \nu} = \Lie_{\phi}~g_{\mu \nu} = \Lie_{\psi}~g_{\mu \nu} = 0,
\ee
where $\Lie$ stands for the Lie derivative with respect to the adequate Killing vector
fields. We shall also assume that the scalar field $\varphi$ is invariant due to 
the action of Killing vector fields, namely
\be
\Lie_{k}~\varphi = \Lie_{\phi}~\varphi  = \Lie_{\psi}~\varphi = 0.
\ee
The metric in the spacetime under consideration can be written 
in the Weyl-Papapetrou form as follows:
\be
ds^2 = - {\rho^2 \over f} dt^2 + f_{ab} \bigg(
dx^{a} + \omega^{a} dt \bigg) \bigg(
dx^{b} + \omega^{b} dt \bigg) + {e^{2 \sigma} \over f}
\bigg( d\rho^2 + dz^2 \bigg),
\label{pap}
\ee
where all functions appearing in the above metric have the only $\rho$ and $z$ dependence.
Furthermore,
the metric (\ref{pap}) can also be rearrange as
\be
ds^2 = \sigma_{ab} dx^{a} dx^{b} + \ga_{ij} dx^{i} dx^{j},
\ee
where $a, b = t, \phi, \psi$ and comprises the first two components in expression (\ref{pap})
while  $i, j = \rho, z$ and describes $\ga_{ij} = X g_{ij}$. $g_{ij}$
stands for the metric of a flat spacetime written in $(\rho, z)$ coordinates.
The conformal factor is equal to 
$X = e^{2 \sigma}/f$.
Using rules of conformal transformation, after some algebra, we find expressions for the 
Ricci tensor components:
\ben
R_{ij} = {}^{(\ga)} R_{ij} &+& {1 \over 2} \ga_{ij} {}^{(\ga)} \na^2 \ln X -
{1 \over 2 \rho X} \bigg(
{}^{(\ga)} \na_{i} \rho~ {}^{(\ga)} \na_{j} X + {}^{(\ga)} \na_{i}X~ {}^{(\ga)} \na_{j} \rho
- \ga_{ij} {}^{(\ga)} \na^{k} \rho~ {}^{(\ga)} \na_{k} X
\bigg)  \\ \nonumber
&-& {1 \over \rho} {}^{(\ga)} \na_{i}{}^{(\ga)} \na_{j} \rho + {1 \over \rho^2}
{}^{(\ga)} \na_{i} \rho~ {}^{(\ga)} \na_{j} \rho +
{1 \over 4} {}^{(\ga)} \na_{i} \sigma^{ab}~ {}^{(\ga)} \na_{j} \sigma_{ab}.
\een
Consequently equations of motion yield
\ben
R_{\rho \rho} - R_{z z} &+& {1 \over \rho X} {}^{(g)}\na_{\rho} X - {1 \over \rho^2}
+ {1 \over 4} \bigg(
{}^{(g)}\na_{z} \sigma^{ab} {}^{(g)}\na_{z} \sigma_{ab} -
{}^{(g)}\na_{\rho} \sigma^{ab} {}^{(g)}\na_{\rho} \sigma_{ab}
\bigg) = {2 \over \rho}~ {}^{(g)} \na_{\rho} \sigma , \\
2 R_{z\rho} &+& {1 \over \rho X} {}^{(g)}\na_{z} X - 
{}^{(g)}\na_{z} \sigma^{ab} {}^{(g)}\na_{\rho} \sigma_{ab} = 
{2 \over \rho}~ {}^{(g)}\na_{z} \sigma, \\
R_{zz} + R_{\rho \rho} &-& {}^{(g)}\na_{m}{}^{(g)}\na^{m} \ln X +
{1 \over \rho X} {}^{(g)}\na_{\rho}X - {1 \over \rho X}
{}^{(g)}\na^{j} \rho {}^{(g)}\na_{j}X - {1 \over \rho^2} + \\ \nonumber
&-& {1 \over 4} \bigg(
{}^{(g)}\na_{\rho} \sigma^{ab}~{}^{(g)}\na_{\rho} \sigma_{ab} +
{}^{(g)}\na_{z} \sigma^{ab}~{}^{(g)}\na_{z} \sigma_{ab}
\bigg) = - 2 {}^{(g)}\na^{j} {}^{(g)}\na_{j} \sigma,
\een
where ${}^{(g)}\na$ is the derivative with respect to $g_{ij}$ metric.
From the above implies directly that $\sigma(\rho, z)$ reduces to the sum of two
components, i.e.,
\be
\sigma = \sigma(vac) + \sigma(\varphi),
\label{ss}
\ee
where $\sigma(vac)$ is the solution of five dimensional vacuum equations of motion
while $\sigma(\varphi)$ is connected with the solution of matter equations.\\
The equations of motion for self-gravitating non-linear $\sigma$-model yield
\be
{1 \over \rho} 
{}^{(g)}\na_{z} \sigma(\varphi) = {1\over 2} G_{AB}(\varphi (x)) 
\bigg( 
{}^{(g)}\na_{\rho}\varphi^{A}~ {}^{(g)}\na_{z}\varphi^{B} + 
{}^{(g)}\na_{z}\varphi^{A}~ {}^{(g)}\na_{\rho}
\varphi^{B} \bigg),
\label{m1}
\ee
\be
{1 \over \rho}
{}^{(g)}\na_{\rho} \sigma(\varphi) = {1\over 2} G_{AB}(\varphi (x)) 
\bigg( 
{}^{(g)}\na_{\rho} \varphi^{A}~ {}^{(g)}\na_{\rho} \varphi^{B} 
- {}^{(g)}\na_{z} \varphi^{A}~
{}^{(g)} \na_{z} \varphi^{B} \bigg),
\label{m2}
\ee
\be
{}^{(g)}\na_{m} {}^{(g)}\na^{m} \sigma (\varphi) = -{1\over 2} G_{AB}(\varphi (x)) 
\bigg( 
{}^{(g)}\na_{\rho} \varphi^{A} {}^{(g)}\na_{\rho} \varphi^{B}
+ {}^{(g)}\na_{z} \varphi^{A}
{}^{(g)}\na_{z} \varphi^{B} \bigg).
\label{m3}
\ee
In order to prove the uniqueness theorem for five-dimensional $\sigma$-model
we shall use the strategy presented in Ref.\cite{heu95}. Namely, we choose a
two-dimensional vector
\be
\Pi_{j} = \rho~ {}^{(g)}\na_{j} e^{- \sigma(\varphi)}.
\ee
Henceforth, by virtue of
Stokes' theorem for $\Pi_{j}$ vector and integration over
the domain of outer communication 
$<<{\cal J}>>$ we find the following:
\ben \label{stok}
D_{\p {\cal J}} &=& 
\int_{\p {\cal J}} \rho e^{- \sigma(\varphi)}
\bigg( 
 {}^{(g)}\na_{z} 
\sigma(\varphi)~ d\rho -  {}^{(g)}\na_{\rho} \sigma(\varphi)~ dz \bigg)
\\ \nonumber
&=&
\int_{<<{\cal J}>>} d\rho dz~ \rho e^{- \sigma(\varphi)}
\bigg[ {}^{(g)}\na^{i}  \sigma(\varphi) {}^{(g)}\na_{i} \sigma(\varphi)
- \bigg( {1 \over \rho} {}^{(g)}\na_{\rho}\sigma(\varphi) +
{}^{(g)}\na^{i} {}^{(g)}\na_{i}\sigma(\varphi) \bigg) \bigg].
\een 
From Eqs.(\ref{m1}),(\ref{m2}) and (\ref{m3}) 
it follows in particular that the second term on the right-hand 
side of (\ref{stok}) is greater or equal to zero. 
It implies that the right-hand side is the 
sum of two non-negative terms.\\
Now, let us calculate the left-hand side of expression (\ref{stok}).
In order to do so 
we introduce an ellipsoidal type of coordinates \cite{car}
given as:
\be
\rho^2 = (\la^2 - c^2)(1 - \mu^2), \qquad z = \la \mu,
\ee
where $\mu = \cos 2 \theta$ is chosen in such a way that the event horizon boundary
occurs for a constant value of $\la = c$.
Two rotation axis segments distinguishing the north and south of the horizon
are given by the respective limit
$\mu = \pm 1$. We take into account
the domain of outer communication $<<{\cal J}>>$ 
as a rectangle, i.e.,
\ben
\p {\cal J}^{(1)} &=& \{ \mu = 1,~ \la = c, \dots, R \},\\ \nonumber
\p {\cal J}^{(2)} &=& \{ \la = c,~ \mu = 1, \dots, -1 \},\\ \nonumber
\p {\cal J}^{(3)} &=& \{ \mu = - 1,~ \la = c, \dots, R \},\\ \nonumber
\p {\cal J}^{(4)} &=& \{ \la = R,~ \mu = - 1, \dots, 1 \}.
\een
Next one rewrites the left-hand side of Eq.(\ref{stok}) in ellipsoidal coordinates
as functions of $\la$ and $\mu$
\ben 
D_{\p {\cal J}} &=& 
\int_{\p {\cal J}} \rho e^{- \sigma(\varphi)}
\bigg( 
 {}^{(g)}\na_{z} 
\sigma(\varphi)~ d\rho -  {}^{(g)}\na_{\rho} \sigma(\varphi)~ dz \bigg) = \\ \nonumber
&=& \int_{\p {\cal J}} e^{- \sigma(\varphi)}
\bigg[
(1 - \mu^2) \sigma(\varphi)_{,\mu}~d\la - (\la^2 - c^2) \sigma(\varphi)_{,\la}~d\mu
\bigg],
\een
where using Eqs.(\ref{m1}) and (\ref{m2}) in ellipsoidal type of
coordinates one reaches to the following
expressions for $\sigma(\varphi)_{,\la}$ and $\sigma(\varphi)_{,\mu}$:
\ben \label{sl}
\sigma(\varphi)_{,\la} =
{G_{AB} (1 - \mu^2) \over 2 (\la^2 - \mu^2 c^2)}
\bigg[
- \la (1 - \mu^2)~ \varphi^{A}_{,\mu} \varphi^{B}_{,\mu} +
\la (\la^2 - c^2)~ \varphi^{A}_{,\la} \varphi^{B}_{,\la}
- 2 \mu (\la^2 - c^2)~ \varphi^{A}_{,\la} \varphi^{B}_{,\mu}
\bigg], \\ \label{sm}
\sigma(\varphi)_{,\mu} =
{G_{AB} (\la^2 - c^2) \over 2 (\la^2 - \mu^2 c^2)}
\bigg[
- \mu (1 - \mu^2)~ \varphi^{A}_{,\mu} \varphi^{B}_{,\mu} +
\mu (\la^2 - c^2)~ \varphi^{A}_{,\la} \varphi^{B}_{,\la}
+ 2 \la (1 - \mu^2)~ \varphi^{A}_{,\la} \varphi^{B}_{,\mu}
\bigg].
\een
Having in mind relations (\ref{sl}) and (\ref{sm}) we can establish
that $\sigma(\varphi)_{,\la},
\sigma(\varphi)_{,\mu}$ and $e^{- \sigma(\varphi)}$ remain finite along the boundaries
$\p {\cal J}^{(i)}$ for $i = 1, 2, 3$ 
and they all
vanish along these parts 
of the boundary. 
It remains to consider
the last part of the boundary. In order to do so one uses
the formula for $\la$ in terms of radial coordinates $r$ \cite{mor04}, as follows:	
\be
\la = {r^2 \over 2} + {a^2 + b^2 \over 4} + {\cal O}\bigg( {1 \over r} \bigg),
\ee
where the parameters $a$ and $b$ are bounded with two independent angular
momentum of five-dimensional Kerr black hole, i.e.,
$J_{\phi} = {\pi m a /4 G}$ and $J_{\psi} = {\pi m b /4 G}$.
The mass of black hole is connected with $m$ parameter by the relation
$M = {3 \pi m /8 G}$.
Hence, it implies
\be
D_{\p {\cal J}^{(4)}} = - \lim_{R \rightarrow \infty} \int_{D_{\p {\cal J}^{(4)}}}
e^{- \sigma(\varphi)} (\la^2 - c^2)~ \sigma_{,\la}~d\mu = 
- {1 \over 2} \int_{0}^{{\pi \over 2}}~
\lim_{r \rightarrow \infty} \bigg(
e^{- \sigma(\varphi)} r^3 \sigma_{,r}~ \sin 2 \theta~ d\theta
+ {\cal O}\bigg( {1 \over r^{n}} \bigg)
\bigg),
\ee
where $n \ge 7$.
Having in mind the asymptotical properties of the derivatives of scalar field $\varphi$
\be
\varphi^{A}_{, \theta} = {\cal O}\bigg( {1 \over r^{3/2}} \bigg),
\qquad
\varphi^{A}_{, r} = {\cal O}\bigg( {1 \over r^{5/2}} \bigg),
\ee
we conclude that
the above entire integral vanishes to the fact that 
$\lim_{r \rightarrow \infty} r^3~ \sigma_{,r} = 0$.\\ 
Hence
${}^{(g)}\na^{i}  \sigma(\varphi) {}^{(g)}\na_{i} \sigma(\varphi)$ and 
${1 \over \rho} {}^{(g)}\na_{\rho}\sigma(\varphi) +
{}^{(g)}\na^{i} {}^{(g)}\na_{i}\sigma(\varphi) $ are equal to zero. 
It occurs that    
$\sigma(\varphi)$
is constant in the considered domain of outer communication, but using the fact that 
$\sigma(\varphi)$ tends to zero as $r \rightarrow \infty$ we get that $\sigma(\varphi) = 0$
which in turn implies that $\varphi$ is constant in the entire domain $<<{\cal J}>>$. Just
from Eq.(\ref{ss}) we have obtained the only $\sigma(vac)$ solution of 
equations of motion. In Ref.\cite{mor04}
uniqueness of the asymptotically flat, stationary five-dimensional black hole solution
being the solution of Einstein vacuum equations with regular
event horizon homeomorphic to $S^3$ and admitting three commutating
Killing vector fields (two spacelike and one timelike) was presented.
Thus, we can assert the main conclusion of our work:\\
Theorem:\\
Let us consider a stationary axisymmetric solution to five-dimensional self-gravitating 
non-linear $\sigma$-models with an asymptotically timelike Killing vector field $k_{\mu}$
and two spacelike Killing vector fields $\phi_{\mu}$ and $\psi_{\mu}$.
The scalar field is invariant under the action of the Killing vector fields. Then, the only
black hole solution with regular rotating event horizon in the asymptotically flat,
strictly stationary domain of outer communication is the five-dimensional Myers-Perry vacuum
Kerr black hole solution with a constant mapping $\varphi$.






\begin{references}
%
\def\cmp#1#2#3{{ Commun. Math. Phys.} {\bf #1}, #2 (#3)}
\def\lmp#1#2#3{{ Lett. Math. Phys.} {\bf #1}, #2 (#3)}
\def\hpa#1#2#3{{ Hell. Phys. Acta} {\bf #1}, #2 (#3)}
\def\grg#1#2#3{{ Gen. Rel. Grav.} {\bf #1}, #2 (#3)}
\def\pr#1#2#3{{ Phys. Rev.} {\bf #1}, #2 (#3)}
\def\prl#1#2#3{{ Phys. Rev. Lett.} {\bf #1}, #2 (#3)}
\def\prd#1#2#3{{ Phys. Rev. D} {\bf #1}, #2 (#3)}
\def\pl#1#2#3{{ Phys. Lett} {\bf #1}, #2 (#3)}
\def\pla#1#2#3{{ Phys. Lett. A} {\bf #1}, #2 (#3)}
\def\plb#1#2#3{{ Phys. Lett. B} {\bf #1}, #2 (#3)}
\def\prep#1#2#3{{ Phys. Reports} {\bf #1}, #2 (#3)}
\def\phys#1#2#3{{ Physica} {\bf #1}, #2 (#3)}
\def\jcp#1#2#3{{ J. Comput. Phys.} {\bf #1}, #2 (#3)}
\def\jmp#1#2#3{{ J. Math. Phys.} {\bf #1}, #2 (#3)}
\def\jpm#1#2#3{{ J. Phys. A: Math. Gen.} {\bf #1}, #2 (#3)}
\def\cpr#1#2#3{{ Computer Phys. Rept.} {\bf #1}, #2 (#3)}
\def\cqg#1#2#3{{ Class. Quantum Grav.} {\bf #1}, #2 (#3)}
\def\cma#1#2#3{{ Computers Math. Applic.} {\bf #1}, #2 (#3)}
\def\mc#1#2#3{{ Math. Compt.} {\bf #1}, #2 (#3)}
\def\apj#1#2#3{{ Astrophys. J.} {\bf #1}, #2 (#3)}
\def\apjs#1#2#3{{ Astrophys. J. Suppl.} {\bf #1}, #2 (#3)}
\def\acta#1#2#3{{ Acta Astronomica} {\bf #1}, #2 (#3)}
\def\apl#1#2#3{{Ann. Physik. (Leipzig)} {\bf #1}, #2 (#3)}
\def\anp#1#2#3{{Ann. Phys. } {\bf #1}, #2 (#3)}
\def\sa#1#2#3{{ Sov. Astro.} {\bf #1}, #2 (#3)}
\def\sia#1#2#3{{ SIAM J. Sci. Statist. Comput.} {\bf #1}, #2 (#3)}
\def\aa#1#2#3{{ Astron. Astrophys.} {\bf #1}, #2 (#3)}
\def\mnras#1#2#3{{ Mon. Not. R. astr. Soc.} {\bf #1}, #2 (#3)}
\def\npb#1#2#3{{ Nucl. Phys. B} {\bf #1}, #2 (#3)}
\def\prsla#1#2#3{{ Proc. R. Soc. London, Ser. A} {\bf #1}, #2 (#3)}
\def\jhep#1#2#3{{ JHEP} {\bf #1}, #2 (#3)}
\def\nuc#1#2#3{{Nuovo Cimento B } {\bf #1}, #2 (#3)}
\def\ijmp#1#2#3{{Int. J. Mod. Phys. D} {\bf #1}, #2 (#3)}
\def\atmp#1#2#3{{Adv. Theor. Math. Phys.} {\bf #1}, #2 (#3)}
\def\ptps#1#2#3{{Prog. Theor. Phys. Suppl.} {\bf #1}, #2 (#3)}
\def\lmp#1#2#3{{Lett. Math. Phys. } {\bf #1}, #2 (#3)}
%
\def\hepph#1#2{{ hep-ph }{\bf #1} (#2)}
\def\hepth#1#2{{ hep-th }{\bf #1} (#2)}
\def\grqc#1#2{{ gr-qc }{\bf #1} (#2)}
%



\bibitem{hor}P.Horava and E.Witten, \npb{460}{506}{1996},\\
P.Horava and E.Witten, \npb{475}{94}{1996}.
\bibitem{gid}S.B.Giddings and S.Thomas, \prd{65}{056010}{2002},\\
S.B.Giddings, {\it Black Hole Production in TeV-scale Gravity, and the Future of High Energy Physics}
\hepph{0110127}{2001},\\
S.B.Giddings, \grg{34}{1775}{2002},\\
D.M.Eardley and S.B.Giddings, \prd{66}{044011}{2002}.
\bibitem{isr}W.Israel, \pr{164}{1776}{1967}.
\bibitem{mil73}H.M\"uller zum Hagen, C.D.Robinson and H.J.Seifert,
\grg{4}{53}{1973},\\
H.M\"uller zum Hagen, C.D.Robinson and H.J.Seifert,
\grg{5}{61}{1974}.
\bibitem{rob77}C.D.Robinson, \grg{8}{695}{1977}. 
\bibitem{bun87}G.L.Bunting G.L and A.K.M.Masood-ul-Alam, 
\grg{19}{147}{1987}.
\bibitem{ru}P.Ruback, \cqg{5}{L155}{1988}. 
\bibitem{ma1}A.K.M.Masood-ul-Alam, \cqg{9}{L53}{1992}.
\bibitem{he1}M.Heusler, \cqg{11}{L49}{1994}.
\bibitem{he93}M.Heusler, \cqg{10}{791}{1993}.
\bibitem{chr99a}P.T.Chru\'sciel, \cqg{16}{661}{1999}.
\bibitem{chr99b}P.T.Chru\'sciel, \cqg{16}{689}{1999}.
\bibitem{car}
B.Carter in {\it Black Holes}, edited
by C.DeWitt and B.S.DeWitt (Gordon and Breach, New York, 1973),\\
B.Carter in {\it Gravitation and Astrophysics}, edited
by B.Carter and J.B.Hartle (Plenum Press, New York, 1987).
\bibitem{rob}
C.D.Robinson, \prl{34}{905}{1975},
\bibitem{maz}P.O.Mazur, \jpm{15}{3173}{1982},\\
P.O.Mazur, \pla{100}{341}{1984},\\
P.O.Mazur, \grg{16}{211}{1984}.
\bibitem{bun}G.L.Bunting, PHD thesis, Univ.of New 
England, Armidale N.S.W., 1983. 
\bibitem{book}P.O.Mazur,
{\it Black Hole Uniqueness Theorems}, \hepth{0101012}{2001},\\
M.Heusler, {\it Black Hole Uniqueness Theorems} 
(Cambridge: Cambridge University Press, 1997).
\bibitem{sta}M.Rogatko, \cqg{14}{2425}{1997},\\
M.Rogatko, \prd{58}{044011}{1998}.
\bibitem{dil}A.K.M.Masood-ul-Alam, \cqg{10}{2649}{1993},\\
M.G\"urses and E.Sermutlu, \cqg{12}{2799}{1995}.
\bibitem{mar01}M.Mars and W.Simon, \atmp{6}{279}{2003}.
\bibitem{rog99}M.Rogatko, \prd{59}{104010}{1999}. 
\bibitem{rog02}M.Rogatko, \cqg{19}{875}{2002}.
\bibitem{gib03}G.W.Gibbons, D.Ida and T.Shiromizu,
\ptps{148}{284}{2003}.
\bibitem{gib02a}
G.W.Gibbons, D.Ida and T.Shiromizu,
\prd{66}{044010}{2002}.
\bibitem{gib02b}
G.W.Gibbons, D.Ida and T.Shiromizu, \prl{89}{041101}{2002}.
\bibitem{kod04}H.Kodama, {\it Uniqueness and Stability of Higher-dimensional Black Holes},
\hepth{0403030}{2004}.
\bibitem{rog03}M.Rogatko, \prd{67}{084025}{2003}.
\bibitem{rog04}M.Rogatko, \prd{70}{044023}{2004}.
\bibitem{mye86}R.C.Myers and M.J.Perry, \anp{172}{304}{1986}.
\bibitem{emp02}R.Emparan and H.S.Reall, \prl{88}{101101}{2002}.
\bibitem{emp04}R.Emparan, \jhep{0403}{064}{2004}.
\bibitem{mor04}Y.Morisawa and D.Ida, \prd{69}{124005}{2004}.
\bibitem{tom04}S.Tomizawa, Y.Uchida and T.Shiromizu, {\it
Twist of Stationary Black Hole/Ring in Five Dimensions},
\grqc{0405134}{2004}.
\bibitem{rog02a}M.Rogatko, \cqg{19}{L151}{2002}.
\bibitem{sup}H.S.Reall, \prd{68}{024024}{2003},\\
J.B.Gutowski, {\it Uniqueness of Five-dimensional Supersymmetric Black Holes},
\hepth{0404079}{2004}.
\bibitem{heu95}M.Heusler, \cqg{12}{2021}{1995}.
\end{references}
\end{document}